\documentclass[12pt]{article}
\usepackage{latexsym}
\usepackage{epsfig}

\newcommand{\beq}{\begin{eqnarray}}
\newcommand{\eeq}{\end{eqnarray}}
\newcommand{\be}{\begin{eqnarray*}}
\newcommand{\ee}{\end{eqnarray*}}

\begin{document}

\centerline{\Large\bf{On the discovery of Birkhoff's theorem}}
\vskip 8mm
\centerline{Nils Voje Johansen}
\vskip 1mm
\centerline{\it Department of Mathematics, University of Oslo, NO-0316 Oslo, Norway.}
\vskip 4mm
\centerline{and}
\bigskip
\centerline{Finn Ravndal}
\vskip 1mm
\centerline{\it Department of Physics, University of Oslo, NO-0316 Oslo, Norway.}

\begin{abstract}

{\footnotesize Birkhoff showed in 1923 that the  Schwarzschild solution for the metric from a point particle was also valid in the 
a priori non-static 
case as long as the spherical symmetry was maintained. This theorem was actually discovered and published two years earlier by an 
unknown Norwegian physicist, J.T. Jebsen. His life and scientific career is briefly chronicled.}
 
\end{abstract}

The static metric $g_{\mu\nu}$ around a point mass $M$ was first found by Schwarzschild\cite{Schwarzschild}. It was derived in 1915 just before
Einstein had completed his general theory and the condition $\det{g_{\mu\nu}} = -1$ had to be imposed, requiring a special choice of coordinates.
Rewriting the result using standard, polar coordinates $(r,\theta,\phi)$ in addition to the time coordinate $t$,  he got the corresponding
line element in the well-known form
\beq
        ds^2 = (1 - 2GM/r)dt^2 - {dr^2\over  1 - 2GM/r} - r^2(d\theta^2 + \sin^2\!\theta d\phi^2)            \label{schwarz}
\eeq
Here $G$ is the gravitational constant and we use $c=1$ units. The derivation we now find in most textbooks can be traced back to 
Hilbert\cite{Hilbert} and Weyl\cite{Weyl} as recently discussed by Antoci\cite{Antoci}. It had also been found around the same time 
by Droste\cite{Droste} who was a student in Leiden, working under the influence of Lorentz\cite{Knox}.

In 1923 Birkhoff in his textbook on modern physics\cite{Birkhoff} showed that the static Schwarz\-schild metric (\ref{schwarz}) is a solution
outside any mass distribution, even when this varies with time as long as the spherical symmetry is maintained. This is now called 
the Birkhoff theorem. A practical and important astrophysical application is the spherical collapse of a star which therefore cannot 
result in any emitted gravitational radiation. From a mathematical point of view, it was derived in an elegant way, concentrating 
on the two essential coordinates $(t,r)$. In this way it can easily be generalized to spacetimes of higher dimensions. An equivalent 
result was derived the same year by Alexandrow using the variational principle\cite{Alexandrow} and shortly later by 
Eiesland\cite{Eiesland}\footnote{John Eiesland was born in 1867 in Norway where he obtained  his secondary education. In 1889 he emigrated 
to the U.S.A.
where got his PhD in mathematics from Johns Hopkins University nine years later. In 1907 he was appointed professor in mathematics at the 
University of West Virginia where he remained until his retirement in 1938. He died in 1950\cite{Breiteig}. A prelimenary version of
his paper in the static case was presented already in 1921\cite{Bull}.}

Recently it has been pointed out by Deser and Franklin\cite{DF} that the theorem was actually discovered three years earlier by Jebsen and
published in 1921 in the proceedings of the Swedish Academy of Sciences\cite{Jebsen}. The derivation is physically motivated, using
the Gaussian coordinates introduced by Hilbert\cite{Hilbert} and Droste\cite{Droste}. But while Hilbert had derived the metric (\ref{schwarz})
from varying the Einstein-Hilbert action, Jebsen derived his more general result directly from Einstein's field equation. Today this is the 
way the theorem is usually proved. The paper was cited for the 
first time in a short abstract in English in 1921\cite{Hodgson} and also mentioned the same year by Oseen in his review of the theory of 
relativity\cite{Oseen}. However, in the following years it was 
essentially completely overlooked except for a reference by Synge in his book on general relativity\cite{Synge}. In addition, nothing seemed 
to be known about Jebsen. Spurred by the apparent Scandinavian background of his contribution, we have now managed to map out most of his 
life and scientific achievements\cite{VJR}.

J\o rg Tofte Jebsen was born in 1888 in Berger, a small town outside Oslo. After his secondary studies and a couple of years abroad, 
he started to study 
physics in 1909 at the University of Oslo. The spring of 1914 he spent at the University of Berlin under the tutelage of Pohl to work on X-rays. 
Returning to Norway, he spent some time helping establish physics at the new, technical university in 
Trondheim. In 1918 he finished his
thesis for the final university exam in Oslo and was apparently so pleased with the result that he asked for it to be evaluated for a doctors
degree. In it he undertook some investigations of electrodynamical problems  within the framework of the special theory of relativity.

In Oslo at that time there was a general interest in Einstein's theories among the students while these new ideas were received with a bit 
more caution by members of the faculty. Until then there had not been any Norwegian scientific publications in this new field. The local faculty
committee which was appointed to evaluate Jebsen's thesis, therefore took contact with Oseen at the University of Uppsala in Sweden. 
He was no great expert either but carried much weight in such matters. Oseen pointed out a few weak points in the thesis and suggested several
improvements. As a result, Jebsen changed his mind about a possible doctors degree  and he was awarded an ordinary university degree in the spring 
of 1919. During this period he had found out that he suffered from tuberculosis and sought treatment. 

However, Oseen must have seen that Jebsen had talent and valuable insight in these new theories. During their discussions
he was invited to come and visit Oseen in Uppsala. In the fall of 1919 he moved there with support from Norway. At the same time 
Oseen started a lecture series on Einstein's general theory, apparently for the first time in Sweden. During the following winter Jebsen worked on
spherical symmetrical solutions of Einstein's gravitational field equation. In the spring of 1920 he had finished the paper {\it \"Uber die
allgemeinen kugelsymmetrischen L\"osungen der Einsteinschen Gravitationsgleichungen im Vakuum} in which he derived the Schwarzschild solution
(\ref{schwarz}) without assuming time-independence. It was sent to the Swedish Academy of Sciences for publication.

Summer of 1920 found Jebsen back in Oslo. At this time Einstein was also there for ten days by invitation from the local 
Student Association\cite{NVJ}. 
He gave three public and very popular lectures on his new theories. We don't know if they met, but Jebsen had informed Oseen that he 
planned to attend. In the meantime Jebsen had been informed that his paper could not be published in Sweden because the Academy
was short of money. He thus started to consider other journals. At the same time he also wrote a long and very insightful article
on non-Euclidean geometry and Einstein's general theory for a newly started  mathematical journal\cite{Mat}. It was the first scientific 
exposition in Norway of Einstein's general theory. In the fall of 1920 he moved to Bolzano in Italy for treatment of his tuberculosis.

Oseen refused to see Jebsen's manuscript unpublished for financial reasons and in October 1920 it was finally accepted by the 
Academy. It appeared in print early in 1921\cite{Jebsen}. Jebsen continued a very extensive and interesting correspondence with Oseen on other 
solutions of Einstein's equations and the question of symmetries. His letters have been kept but not Oseen's responses\cite{CHS}. While Jebsen was  
enthusiastic about the general theory, it was known that Oseen was more sceptical, in part because it did not throw any
new light on the structure of the electron\cite{Oseen}.

In Italy Jebsen did not get better from his illness, but still finished a book on Galilei. He died there early in 1922. The same year Oseen was
elected to the Nobel committee for physics. Here he was influential in the award of the 1922 Nobel prize to Bohr for the structure of atoms 
while Einstein at the same occasion was awarded the 1921 prize for the photoelectric effect.

\vskip 2mm

\leftline{\bf Acknowledgement:}

We want to thank Professor Stanley Deser for informing us about Jebsen's  unknown contribution and encouraging us to find out who he actually 
was. In addition, we want to thank Professor Salvatore Antoci for several useful comments.

\end{document}